\documentclass{article}%
\usepackage{amsfonts}
\usepackage{amsmath}%
\usepackage{amssymb}%
\usepackage{graphicx}
\providecommand{\U}[1]{\protect\rule{.1in}{.1in}}

\begin{document}

\title{New applications of the Lambert and generalized Lambert functions to
ferromagnetism and quantum mechanics}
\author{Victor Barsan\\IFIN-HH, Str. Reactorului nr. 30, Magurele 077125, Romania}
\maketitle

\begin{abstract}
The applications of the recent results obtained in the theory of generalized
Lambert functions, to the mean field theory of ferromagnetism and quantum
mechanics are presented. As a consequence, all the predictions of the Weiss
theory of ferromagnetism can be explicitly and exactly formulated. In several
quantum mechanical problems involving delta function potentials, the solutions
of the transcendental eigenvalue equations are expressed in terms of two
parameters generalized Lambert functions. Some others, till unnoticed examples
of eigenvalue equations whose solutions can be written using the Lambert $W$
function, are also presented.

\end{abstract}

\section{\textbf{Introduction}}

In a series of recent papers, Mez\"{o} and co-workers \cite{1}, \cite{2},
\cite{3} and Mugnaini \cite{4} published very interesting results concerning
the generalized Lambert functions. Similar to the Lambert function $W\left(
a\right)  ,$ which is defined as the solution of the transcendental equation
\cite{5}:%

\begin{equation}
xe^{x}=a\label{1}%
\end{equation}
the generalized Lambert function $W\left(  t_{1},t_{2},...t_{n};s_{1}%
,s_{2},...s_{m};a\right)  $ is the solution of a similar, but more complicated
transcendental equation \cite{1}, \cite{4}:%

\begin{equation}
e^{x}\frac{\left(  x-t_{1}\right)  ...\left(  x-t_{n}\right)  }{\left(
x-s_{1}\right)  ...\left(  x-s_{m}\right)  }=a\label{2}%
\end{equation}

If the denominator, respectively the nominator of the fraction in eq.
(\ref{2}) is equal to $1,$ we shall denote the solutions $W\left(
-;s_{1},s_{2},...s_{m};a\right)  ,$ respectively $W\left(  t_{1}%
,t_{2},...t_{n};-;a\right)  .$ In several particular cases of generalized
Lambert functions, explicit series expansions formulas are obtained \cite{1},
\cite{4}. They are very important for explicitly solving some transcendental
equations of magnetism, quantum mechanics, hydrodynamics and of other branches
of physics. The applications in pure mathematics are also very rich.

In the present paper, we shall make use of the solution of the equation%

\begin{equation}
e^{x}\frac{x-t}{x-s}=a\label{3}%
\end{equation}
which is denoted $W\left(  t;s;a\right)  ,$ and of the equation%

\begin{equation}
e^{x}\left(  x-t_{1}\right)  \left(  x-t_{2}\right)  =a\label{4}%
\end{equation}
which is denoted $W\left(  t_{1},t_{2};-;a\right)  .$ The series expansion of
$W\left(  t;s;a\right)  $ near the origin has the form:%

\begin{equation}
W\left(  t;s;a\right)  =t-T\sum_{n=1}^{\infty}\frac{L_{n}^{\prime}\left(
nT\right)  }{n}e^{-nt}a^{n}\label{5}%
\end{equation}
where $T=t-s\neq0,$ and $L_{n}^{\prime}$ is the first derivative of the $n-
$th order Laguerre polynomial.

$W\left(  t;s;a\right)  $ can also be expressed in terms of the $r-$Lambert
function, defined by Mez\"{o} and Baricz \cite{1} as the inverse of the
function $xe^{x}+rx,$ with $r$ - a fixed real number, denoted by $W_{r}.$
Clearly, if $r=0,$ $W_{r}$ becomes the $W$ Lambert function. The connection
between $W\left(  t;s;a\right)  $ and $W_{r}$ can be easily obtained as
(Theorem 3 of \cite{1}):%

\begin{equation}
W\left(  t;s;a\right)  =t+W_{-a\exp\left(  -t\right)  }\left(  a\left(
t-s\right)  e^{-t}\right) \label{6}%
\end{equation}

As noticed by Mez\"{o} and Baricz \cite{1}, $W\left(  t_{1},t_{2}%
,...t_{n};s_{1},s_{2},...s_{m};a\right)  $ is not, properly speaking, a
generalization of $W,$ in the sense that there is no particular choice of the
parameters $t_{1},t_{2},...t_{n};\ s_{1},s_{2},...s_{m}$ or $a, $ which can
transform the generalized Lambert function into the $W$ Lambert function.
However, as just mentioned, the $r-$Lambert function,$\ W_{r},$ has this property.

By inspection of the plots of the functions entering in eq. (\ref{3}), it is
easy to see that $W_{r}$ can have one, two or three branches. Mez\"{o} and
Baricz \cite{1} described rigorously the branch structure of $W_{r}$ and
pointed out that, among these branches, a special role is played by
$W_{1/e^{2}}\left(  x\right)  .\ $For $x=-4/e^{2},\ W_{1/e^{2}}\left(
x\right)  $ is continuos (as everywhere on the real line) but is not
differentiable, and%

\begin{equation}
W_{1/e^{2}}\left(  -4/e^{2}\right)  =-2\label{7}%
\end{equation}

The formula (\ref{6}) must be used with a certain caution, as in the
definition of $W_{r}$, $r$ was considered to be a fixed real number. Under
this condition, the branch structure of $W_{r},$ its derivative and primitive
were obtained. However, in (\ref{6}), $t,\ s$ and $a$ are arbitrary real
parameters, so $r$ is a real function.

Mez\"{o} and Baricz \cite{1} also obtained the series expansion of
$W_{r}\left(  x\right)  $\ around $x=0.$ The coefficients of the series are
expressed in terms of Mez\"{o}-Baricz polynomials $M_{k}^{\left(  n\right)  }$
(\cite{1}, eq. (14)):%

\begin{equation}
W_{r}\left(  x\right)  =\frac{x}{r+1}+\sum_{n=2}^{\infty}M_{n-1}^{\left(
n\right)  }\left(  \frac{1}{r+1}\right)  \frac{x^{n}}{\left(  r+1\right)
^{n}n!}\label{8}%
\end{equation}

Let us refer now to the function $W\left(  t_{1},t_{2};-;a\right)  .$\ Its
series expansion around $a=0$ is given by:%

\begin{equation}
W\left(  t_{1},t_{2};-;a\right)  =t_{1}-\sum_{n=1}^{\infty}\frac{1}%
{n!n}\left(  \frac{ane^{-t_{1}}}{t_{2}-t_{1}}\right)  ^{n}B_{n-1}\left(
-\frac{2}{n\left(  t_{2}-t_{1}\right)  }\right) \label{9}%
\end{equation}
where $B_{n}$ are Bessel polynomials.

The main goal of this paper is to reveal several applications of generalized
Lambert functions $W\left(  t;s;a\right)  $ and\ $W\left(  t_{1}%
,t_{2};-;a\right)  $ in physics, mainly in ferromagnetism and quantum
mechanics, not yet reported in literature. We shall also discuss some other
transcendental equations, which can be solved with the Lambert $W$ function,
but this possibility was not yet noticed, to the best of our knowledge.

\section{Solutions of some transcendental equations using Lambert and
generalized Lambert functions}

We shall study the applications in ferromagnetism and quantum mechanics of two
transcendental equations:%

\begin{equation}
\tanh x=bx\label{10}%
\end{equation}
which is sometimes called Weiss equation of ferromagnetism, and%

\begin{equation}
\tanh x=\frac{b}{x}\label{11}%
\end{equation}
We shall call eq. (\ref{11}) the 'inverse Weiss equation', just in order to
have a convenient terminology in this paper. The equations (\ref{10}),
(\ref{11}) were analyzed by Mez\"{o} and Keady in the context of
hydrodynamics: they describe the phase velocity, respectively the dispersion
of water waves (\cite{3}, eqs. (7) and (5)). Here, we shall discuss the
relevance of these equations in the context of ferromagnetism and quantum mechanics.

Weiss equation can be put in the form:%

\begin{equation}
e^{2x}\frac{x-\frac{1}{b}}{x+\frac{1}{b}}=-1\label{12}%
\end{equation}
and has the solution:%

\begin{equation}
x=\frac{1}{2}W\left(  \frac{2}{b};-\frac{2}{b};-1\right)  =\frac{1}{b}%
+\frac{1}{2}W_{\exp\left(  -2/b\right)  }\left(  -\frac{4}{b}e^{-2/b}\right)
\label{13}%
\end{equation}
which can be easily obtained following the steps described in subsection 2.3.2
of \cite{3} and applying the formula (\ref{6}). Similarly, using the approach
of subsection 2.3.1 of the same paper, applied to the inverse Weiss equation,
we find that it can be put in the form:%

\begin{equation}
e^{2x}\frac{x-b}{x+b}=1\label{14}%
\end{equation}
and has the solution:%

\begin{equation}
x=\frac{1}{2}W\left(  2b;-2b;1\right)  =b+\frac{1}{2}W_{-\exp\left(
-2b\right)  }\left(  4be^{-2b}\right) \label{15}%
\end{equation}

It is also useful to mention that the solution of the equation:%

\begin{equation}
e^{-cx}=a_{0}\left(  x-r\right) \label{16}%
\end{equation}
can be expressed in terms of the Lambert $W$ function \cite{5}:%

\begin{equation}
x=r+\frac{1}{c}W\left(  \frac{ce^{-cr}}{a_{0}}\right) \label{17}%
\end{equation}

\section{An outline of Weiss theory of ferromagnetism}

As Weiss theory of ferromagnetism is one of the main beneficiaries of the
progress made in understanding the explicit form of generalized Lambert
function, we shall outline it in this section, following \cite{6}. A system of
$N$ particles of spin $S$ can be in a ferromagnetic state, described by the
equation of state:%

\begin{equation}
M=M_{0}B_{S}\left(  \frac{\overline{\mu}S}{kT}\left(  H+\lambda M\right)
\right)  ,\ \overline{\mu}=g\mu_{B}\label{18}%
\end{equation}

The magnetization $M$ depends of temperature $T$\ and of external magnetic
field $H$:%

\begin{equation}
M=M\left(  T,H\right) \label{19}%
\end{equation}
The saturation magnetization $M_{0}$\ is:%

\begin{equation}
M_{0}=M\left(  T=0,H=0\right)  =N\overline{\mu}S=NSg\mu_{B}.\label{20}%
\end{equation}

The parameter $\lambda$ represents the molecular field parameter, introduced
by Pierre Weiss in 1907, and $B_{S}$ is the Brillouin function:%

\begin{equation}
B_{S}\left(  x\right)  =\frac{2S+1}{2S}\coth\left(  \frac{2S+1}{2S}x\right)
-\frac{1}{2S}\coth\left(  \frac{1}{2S}x\right) \label{21}%
\end{equation}
If $S=1/2:$%

\begin{equation}
B_{1/2}\left(  x\right)  =\tanh x\label{22}%
\end{equation}

In this paper we shall only discuss the case $S=1/2.$ However, the case
$S=\infty,$ involving the Langevin function $L\left(  x\right)  $, $B_{\infty
}\left(  x\right)  =L\left(  x\right)  ,$ involves essentially the same
mathematics (\cite{3}, subsection 2.2.1). Coming back to the case $S=1/2$, let
us notice that, if the external field is zero, $H=0,$ eq. (\ref{18}) becomes:%

\begin{equation}
M=M_{0}\tanh\left(  \frac{\overline{\mu}S}{kT}\lambda M\right) \label{23}%
\end{equation}
This equation has a non-zero solution, and, equivalently, the system has a
spontaneous magnetization $M\neq0,$ if the temperature $T$ is under a critical
value, namely under the critical temperature $T_{c},$ given by the following relation:%

\begin{equation}
T_{c}=\lambda\frac{N\overline{\mu}^{2}}{4k}\label{24}%
\end{equation}

It is convenient to introduce the reduced parameters:%

\begin{equation}
t_{0}=\frac{T}{T_{c}},\ h=\frac{\overline{\mu}H}{2kT_{c}},\ \ m=\frac{M}%
{M_{0}}\label{25}%
\end{equation}
as the equation of state (\ref{18}) can be written now in a simpler form:%

\begin{equation}
m\left(  t_{0},h\right)  =\tanh\frac{m\left(  t_{0},h\right)  +h}{t_{0}%
}\label{26}%
\end{equation}
We choose the notation $t_{0}$ instead of the usual notation $t$ for the
reduced temperature, in order to avoid any confusion with the parameter $t$
entering in the formulas concerning the generalized Lambert function $W\left(
t;s;a\right)  $. The reduced magnetization in zero external field is:%

\begin{equation}
m\left(  t_{0},0\right)  =\tanh\frac{m\left(  t_{0},0\right)  }{t_{0}%
}\label{27}%
\end{equation}
and the critical isotherm is given by the equation:%

\begin{equation}
m\left(  1,h\right)  =\tanh\left(  m\left(  1,h\right)  +h\right) \label{28}%
\end{equation}

It can be also written as:%

\begin{equation}
h=\tanh^{-1}m\left(  1,h\right)  -m\left(  1,h\right) \label{29}%
\end{equation}
In the critical region, $m,$ $h\ll1$; as\newline%
\begin{equation}
\tanh^{-1}x=\frac{1}{2}\ln\frac{1+x}{1-x}=\allowbreak x+\frac{1}{3}%
x^{3}+O\left(  x^{5}\right)  ,\ x\ll1\label{30}%
\end{equation}
we get:%

\begin{equation}
h\simeq\frac{1}{3}m\left(  1,h\right)  ^{3},\ \ h,m\ll1\label{31}%
\end{equation}

Similarly, in the absence of the magnetic field, near $t_{0}=1,$ putting
$t_{0}=1-\varepsilon,$ (\ref{27}) gives:%

\begin{equation}
m=\tanh\frac{m}{t_{0}}=\tanh\frac{m}{1-\varepsilon}\simeq\tanh m\left(
1+\varepsilon\right)  \simeq m\left(  1+\varepsilon\right)  -\frac{1}{3}%
m^{3}\left(  1+\varepsilon\right)  ^{3}\label{32}%
\end{equation}
or:%

\begin{equation}
1-t_{0}\simeq\frac{1}{3}m^{2},\ \ \ \ m,1-t_{0}\ll1\label{33}%
\end{equation}
The behavior of the reduced spontaneous magnetization near the reduced
critical temperature $t_{c}=1$%

\begin{equation}
m\sim\sqrt{1-t_{0}}\label{34}%
\end{equation}
is typical for a mean field theory of critical phenomena \cite{6}. Actually,
the formulas (\ref{31}), (\ref{33}), (\ref{34}) give the critical behavior of
the Weiss ferromagnet. As the spontaneous magnetization decreases
monotonically when temperature increases, its maximum value is reached at
$T=0,$ and:%

\begin{equation}
m\left(  0,0\right)  =1\label{35}%
\end{equation}

\section{Applications of the new results concerning generalized Lambert
functions to ferromagnetism}

Replacing in eq. (\ref{10})%

\begin{equation}
x\rightarrow\frac{m\left(  t_{0}\right)  }{t_{0}},\ b\rightarrow
t_{0}\label{36}%
\end{equation}
and using the formula (\ref{6}), we get the following relations for the
reduced magnetization in the absence of the magnetic field:%

\begin{equation}
m\left(  t_{0}\right)  =\frac{t_{0}}{2}W\left(  \frac{2}{t_{0}};-\frac
{2}{t_{0}};-1\right)  =1-2\sum_{n=1}^{\infty}\frac{L_{n}^{\prime}\left(
4n/t_{0}\right)  }{n}\left(  -e^{-2/t_{0}}\right)  ^{n}\label{37}%
\end{equation}
Actually, eq. (\ref{37}) can be directly obtained from \cite{3}, eq. (7),
making the replacement $y_{c}\rightarrow t_{0}.$ It is easy to see that the condition%

\begin{equation}
m\left(  0\right)  =1\label{38}%
\end{equation}
is fulfilled, as $L_{1}^{\prime}\left(  x\right)  =-1$ and $\lim
_{t_{0}\rightarrow0}e^{-2/t_{0}}=0.$ Due to the exponential term, the series
in the r.h.s of (\ref{37}) is rapidly convergent. The expression (\ref{37}) of
the magnetization can be also expressed in terms of the $r-$Lambert function
as (see eq. (\ref{6})):%

\begin{equation}
W\left(  \frac{2}{t_{0}};-\frac{2}{t_{0}};-1\right)  =\frac{2}{t_{0}}%
+W_{\exp\left(  -2/t_{0}\right)  }\left(  -\frac{4}{t_{0}}\exp\left(
-2/t_{0}\right)  \right) \label{39}%
\end{equation}
In the case of critical temperature, $t_{0}=1$ and the index of $W_{r}$ takes
its critical value, namely:%

\begin{equation}
r=\frac{1}{e^{2}}\label{40}%
\end{equation}
In this case:%

\begin{equation}
W\left(  2;-2;-1\right)  =2+W_{1/e^{2}}\left(  -\frac{4}{e^{2}}\right)
=0\label{41}%
\end{equation}
according to eq. (\ref{7}). Consequently, according to (\ref{41}), the reduced
magnetization at the critical temperature is zero:%

\begin{equation}
m\left(  1\right)  =0\label{42}%
\end{equation}

More than this, as mentioned just before eq. (\ref{6}), the magnetization is
not differentiable in $t=1,$ but is still continuos. This behavior is
compatible with the aspect of the experimental curve of reduced spontaneous
magnetization at critical temperature (see Fig. 1).

\begin{figure}
\begin{center}
\includegraphics[width=\textwidth]{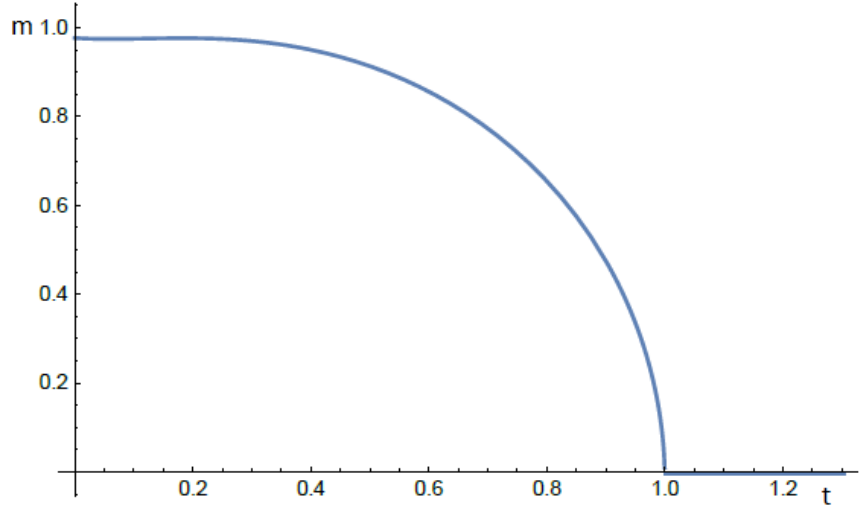}
\end{center}
\caption{The reduced spontaneous magnetization of the Weiss model.}
\end{figure}

If the magnetic field is non-zero, eq. (\ref{26}) can be written as:%

\begin{equation}
\tanh\left(  \frac{m\left(  t_{0},h\right)  }{t_{0}}+h\right)  =t_{0}\left(
\frac{m\left(  t_{0},h\right)  }{t_{0}}+h\right)  -t_{0}h\label{43}%
\end{equation}
or, putting%

\begin{equation}
u=\frac{m\left(  t_{0},h\right)  }{t_{0}}+h\label{44}%
\end{equation}
as:%

\begin{equation}
\tanh u=t_{0}\left(  u-h\right) \label{45}%
\end{equation}
or, equivalently:%

\begin{equation}
e^{2u}\frac{u-h-\frac{1}{t_{0}}}{u-h+\frac{1}{t_{0}}}=-1\label{46}%
\end{equation}

According to (\ref{13}), the solution of this equation is:%

\begin{equation}
u=\frac{1}{2}W\left(  2h+\frac{2}{t_{0}};2h-\frac{2}{t_{0}};-1\right)
\label{47}%
\end{equation}
so, finally we get:%

\begin{equation}
m\left(  t_{0},h\right)  =\frac{t_{0}}{2}W\left(  2h+\frac{2}{t_{0}}%
;2h-\frac{2}{t_{0}};-1\right)  -h\label{48}%
\end{equation}
or, using eq. (\ref{6}):%

\begin{equation}
m\left(  t_{0},h\right)  =W_{\exp\left(  -2h-2/t_{0}\right)  }\left(  \frac
{4}{t_{0}}e^{-2h-2/t_{0}}\right)  -h\left(  1-t_{0}\right)  +1\label{49}%
\end{equation}

The critical isotherm is obtained making in the previous formula the
replacement $t_{0}\rightarrow1:$%

\begin{equation}
m\left(  1,h\right)  =W_{\exp\left(  -2\left(  h-1\right)  \right)  }\left(
4e^{-2\left(  h-1\right)  }\right)  +1\label{50}%
\end{equation}
Let us notice that the $r-$Lambert functions entering in eqs. (\ref{39}),
(\ref{49}), (\ref{50}) have quite special forms:%

\begin{equation}
W_{x}\left(  2x\ln x\right)  ,\ W_{x}\left(  2\left(  x-a\right)  \ln
x\right)  ,\ W_{x}\left(  bx\right) \label{51}%
\end{equation}

These results are important, as they allow explicit calculation of any
physical quantity of Weiss theory of ferromagnetism, for $S=1/2.$ There are
very few realistic physical theories in such a situation - evidently, a
consequence of the fact that the Brillouin function $B_{1/2}$ can be inverted
using generalized Lambert functions. We can wonder if there is another
Brillouin function with this property.

The answer can be easily obtained, if we write the Brillouin function
$B_{S}\left(  x\right)  $\ as a ratio of two polynomials in $\exp\left(
x/2S\right)  $: there is no other $B_{S},$ with finite $S,$ which can be
inverted using the generalized Lambert functions. Only for $S=\infty,$ the
Langevin function, $L\left(  x\right)  $, can be inverted, as previously
noted. It has a somewhat more complicated form, compared with (\ref{51}),
namely (see subsection 2.2.1 of \cite{3} and eq. (\ref{6}) of this paper):%

\begin{equation}
L^{-1}\left(  a\right)  =-\frac{4}{a+1}-2W_{-\frac{a-1}{a+1}\exp\left(
-\frac{2}{a+1}\right)  }\left(  -\frac{4}{\left(  a+1\right)  ^{2}}\exp\left(
-\frac{2}{a+1}\right)  \right) \label{52}%
\end{equation}

So, the $W_{r}$ function in (\ref{52}) has the form $W_{x}\left(  bx\right)  $
(see eq. (\ref{51})), with $b=4/\left(  a^{2}-1\right)  .$

One of the tasks of the theory is to show that eq. (\ref{49}) is compatible
with the critical behavior of a Weiss ferromagnet, eqs. (\ref{31}),
(\ref{34}). The remark inserted just after eq. (\ref{42}) is a step in this direction.

\section{Applications of Lambert and generalized Lambert functions to quantum
mechanics}

The first appearance of a Lambert equation in quantum mechanics is probably a
toy model for the ionized $H_{2}$ molecule: an electron moving in 1D, in the
potential of two identical, attractive delta functions, simulating the two
hydrogen nuclei (protons). The model was proposed by Frost in 1956 \cite{7}.
Its eigenvalue equation is a transcendental equation in $c:$%

\begin{equation}
c=g\left(  1\pm e^{-cR}\right) \label{53}%
\end{equation}
with the solutions:%

\begin{equation}
c^{\left(  \pm\right)  }=g+\frac{1}{R}W\left(  \pm Rg\cdot e^{-Rg}\right)
\label{54}%
\end{equation}
which give the electron eigenenergies:%

\begin{equation}
E^{\left(  \pm\right)  }=-\frac{1}{2}\left(  c^{\left(  \pm\right)  }\right)
^{2}\label{55}%
\end{equation}
The 'plus' sign corresponds to the ground state, and the 'minus' - to the
(unique) excited state. The results (54), (55) were obtained in 1993 by Scott
et al. \cite{8}. We repeated this short calculation for reasons of completeness.

It is quite surprising that, since 1993 till now, the eq. (\ref{53}) was
discussed in a large number of papers, the last one being \cite{9} (see eq.
(16) of this paper), but its connection with the Lambert function remained
unnoticed; apparently, the only exception is provided by \cite{1}. Anyway, we
must keep in mind that the spectrum of a particle moving in 1D in the
potential of two symmetric, attractive delta functions can be expressed in
terms of the Lambert function.

Actually, the conclusion is more general: the Lambert and the generalized
Lambert functions systematically appear in spectral problems of Schroedinger
equations with delta function potentials. We shall illustrate this conclusion
with several examples.

An attractive delta function situated near an impenetrable wall (see for
instance \cite{10}, problem 3.9)%

\begin{equation}
-\frac{\hbar^{2}g}{2m}\delta\left(  x-a\right) \label{56}%
\end{equation}
leads to a bound state energy condition:%

\begin{equation}
\tanh k=\frac{k}{g-k},\label{57}%
\end{equation}
with the solution:%

\begin{equation}
k=-\frac{1}{2}\left(  g+W\left(  ge^{g}\right)  \right) \label{58}%
\end{equation}

An attractive delta potential:%

\begin{equation}
-\frac{\hbar^{2}g}{2m}\delta\left(  x-a\right) \label{59}%
\end{equation}
in the middle of an infinite square well leads to (\cite{10},
problem 3.11):%

\begin{equation}
\tan kL=\frac{2}{g^{2}L}kL\label{60}%
\end{equation}

The case of two different delta functions (\cite{10}, problem 3.11)%

\begin{equation}
V\left(  x\right)  =\frac{\hbar^{2}g_{1}}{2m}\delta\left(  x+a\right)
+\frac{\hbar^{2}g_{2}}{2m}\delta\left(  x-a\right) \label{61}%
\end{equation}
produces a richer physics. The attractive, symmetric case $\left(  g_{1}%
=g_{2}\right)  $, firstly analyzed by Frost \cite{7}, was already discussed
(see eq. (\ref{55}) of this paper).

The case of antisymmetric potentials $\left(  g_{1}=-g_{2}=g\right)  $ leads
to the following eigenvalue equation (\cite{10}, Problem 3.12.b):%

\begin{equation}
e^{K}=\frac{4a^{2}g^{4}}{\left(  2ag^{2}-K\right)  \left(  2ag^{2}+K\right)
}\label{62}%
\end{equation}
with the solution given by a generalized Lambert function:%

\begin{equation}
K=W\left(  2ag^{2},-2ag^{2};-;-4a^{2}g^{4}\right) \label{63}%
\end{equation}

The case of attractive but different delta functions was solved by Mez\"{o}
and Keady \cite{1}.

Even a single attractive delta function can produce a spectrum given by a
generalized Lambert function - namely by $W\left(  t;s;x\right)  $ - if the
boundary conditions are less usual \cite{11}. The Dirichlet boundary
conditions generate an eigenvalue equation for the wave vector $k$ similar to
the Weiss equation:%

\begin{equation}
\tanh ka=\frac{k}{\Omega}\label{64}%
\end{equation}
($a,\ \Omega$ are parameters) and the Neumann boundary
conditions, one similar to the inverse Weiss equation:%

\begin{equation}
\tanh ka=\frac{\Omega}{k}\label{65}%
\end{equation}

Going further to higher dimensions, let us notice that the rotational spectrum
of an attractive delta-shell potential (see \cite{12}, eq. (13)) is given by
an equation:%

\begin{equation}
k\left(  1+\coth kr\right)  =V_{0}\label{66}%
\end{equation}
which can be solved using the Lambert function:%

\begin{equation}
k=\frac{V_{0}}{2}+\frac{1}{2r}W\left(  -rV_{0}\exp\left(  -rV_{0}\right)
\right) \label{67}%
\end{equation}

The most general non-relativistic quantum mechanical problems involving
$\delta$ and $\delta^{\prime}$ functions were studied from the perspective of
mathematical physics by Albeverio \cite{13}. Finitely many point interactions
in 3D might lead to eigenvalues equations like (see \cite{13}, eq. (1.1.81))%

\begin{equation}
y=\pm e^{-y}-4\pi\alpha L\label{68}%
\end{equation}
with the solutions:%

\begin{equation}
y^{\left(  \pm\right)  }=-4\pi\alpha L+W\left(  e^{\pm4\pi\alpha L}\right)
\label{69}%
\end{equation}

Both periodic delta interactions and infinitely many $\delta^{\prime}$
interactions in 1D might produce Weiss and inverse Weiss equations (see
\cite{13}, eq. (2.3.25), (3.39), (3.40)). The physical applications cover toy
models for crystals with impurities, Kronig - Penney models etc. Other similar
applications are discussed by Valluri \cite{14}.

\section{Conclusions}

Due to the interesting results obtained recently in the study of the
generalized Lambert equation, the exact analytical solution of Weiss equation
can be explicitly written, as a series expansion. In this way, the Weiss
theory of ferromagnetism becomes one of the very few cases when all the
predictions of a quite realistic description of a physical system can be
exactly and analytically obtained. The applications of the explicit formulas
obtained for the generalized Lambert functions $W$ $\left(  t;s;a\right)  $
and $W$ $\left(  t_{1},t_{2};-;a\right)  $\ to ferromagnetism and quantum
mechanics are also discussed. Finally, the author draws attention upon several
transcendental eigenvalues equations of quantum mechanics, whose solutions can
be expressed in terms of the $W$ Lambert function.

The author acknowledges the financial support of the IFIN-HH - ANCSI project
PN 16 42 01 01/2016 and to the IFIN-HH - JINR Dubna grant 04-4-1121-2015/17.

\end{document}